\documentclass[aps,prb,twocolumn,superscriptaddress,showpacs]{revtex4-1}

\usepackage{amsmath,amssymb,graphics,epsfig,epstopdf,color,verbatim,ulem,braket,tabularx}
\usepackage[colorlinks,linkcolor=blue,citecolor=blue,urlcolor=blue]{hyperref}

\begin{document}

\title{Quantum Monte Carlo study of strange correlator in interacting topological insulators}
\author{Han-Qing Wu}
\author{Yuan-Yao He}
\address{Department of Physics, Renmin University of China, Beijing 100872, China}
\author{Yi-Zhuang You}
\affiliation{Department of Physics, University of California,
Santa Barbara, California 93106, USA}
\author{Cenke Xu}
\affiliation{Department of Physics, University of California,
Santa Barbara, California 93106, USA}
\author{Zi Yang Meng}
\affiliation{Beijing National Laboratory
for Condensed Matter Physics, and Institute of Physics, Chinese
Academy of Sciences, Beijing 100190, China}
\author{Zhong-Yi Lu}
\address{Department of Physics, Renmin University of China, Beijing 100872, China}

\begin{abstract}
  Distinguishing the nontrivial symmetry-protected topological (SPT)
  phase from the trivial insulator phase in the presence of electron-electron
  interaction is an urgent question to the study of topological
  insulators, due to the fact that most of the topological indices
  defined for free electron systems are very likely unsuitable for
  interacting cases. In this work, we demonstrate that the strange
  correlator is a sensitive diagnosis to detect SPT states in
  interacting systems. Employing large-scale quantum Monte Carlo
  (QMC) simulations, we investigate the interaction-driven quantum
  phase transition in the Kane-Mele-Hubbard model. The transition
  from the quantum spin Hall insulator at weak interaction to an antiferromagnetic
  Mott insulator at strong interaction can be readily detected by
  the momentum space behavior of the strange correlator in single-particle,
  spin, and pairing sectors. The interaction effects on the symmetry-protected
  edge states in various sectors, i.e., the helical Luttinger
  liquid behavior, are well captured in the QMC measurements of strange
  correlators. Moreover, we demonstrate that the strange correlator is
  technically easier to implement in QMC and more robust in performance
  than other proposed numerical diagnoses for interacting topological states,
  as only static correlations are needed. The attempt in this work paves
  the way for using the strange correlator to study interaction-driven
  topological phase transitions in fermionic as well as bosonic systems.
\end{abstract}


\date{\today} \maketitle

\section{INTRODUCTION}

Topological insulators (TIs) are usually
defined as systems with bulk spectra similar to those of trivial
insulators but with nontrivial, $i.e.$ gapless or degenerate, boundary
spectrums when and only when the systems (including the boundaries)
preserve certain symmetries. By now the noninteracting TIs have been
fully classified and understood, for example, as shown in
Refs.~\onlinecite{Ludwig_Class1,Ludwig_Class2,Kitaev_Class}.
Besides the boundary states, which are experimentally most
relevant, the noninteracting TIs can also be characterized by a
topological index defined for the bulk band structure, namely,
even if a TI has the similar bulk spectrum as a trivial insulator,
it does have a very different ground-state wave function which is
characterized by the topological indices, for example, the Thouless¨CKohmoto¨CNightingale¨Cden Nijs (TKNN)
number for the integer quantum Hall state~\cite{TKNN1982} and the
$Z_2$ index for the quantum spin Hall insulator~\cite{Kane2005a,
Kane2005b}. To generalize the notion of topological
insulator to interacting systems, symmetry-protected
topological (SPT) order~\cite{Wen_SPT, Wen_SPT2} was proposed for the ground
states of many-body quantum systems that have a symmetry
and a finite energy gap with short range quantum
entanglement. So far most of the techniques and topological
indices introduced for noninteracting topological insulators are
very likely unsuitable for interacting cases, since in many cases
interaction can change (or reduce) the classification of
topological insulators~\cite{Fidkowski1, Fidkowski2, XLQi2013,
HYao2013, SCZhang2012, ZCGu2014, Fidkowski2013, Senthil2014,
YZYou2014_TSC, YZYou2014_SPT, YZYou2014_ITI}. Thus a more general
technique to identify interacting TIs (or SPT states) based on
their bulk wave functions is urgently demanded for studying
topological insulators in the presence of electron-electron interactions.

In principle, given a bulk wave function, we can always compute
its entanglement spectrum and use it as a diagnosis for
interacting TI~\cite{Haldane2008}. However, this technique is
numerically challenging. For strongly correlated electron systems,
in one dimension (1D) we are able to obtain the bulk wave function
and entanglement spectrum from exact diagonalization (ED) and
density matrix renormalization group (DMRG) calculations, but in
two dimensions (2D) and higher, it is very difficult to obtain the
bulk wave functions for interacting systems simply because the
dimension of Hilbert space increases exponentially with the number
of electrons. In 2D, there has been recent progress by employing
quantum Monte Carlo simulations to access the entanglement
spectrum~\cite{Grover2013, Assaad2014, Assaad2015}, but the
approach is arduous, as one needs to first bifurcate the already
small finite-size system (the simulation efforts of QMC scale
polynomially with system size to high power) and then perform
analytical continuation to obtain the real-frequency entanglement
spectrum from the reduced density matrix in imaginary
time~\cite{Assaad2014, Assaad2015}. The analytical continuation
~\cite{Jarrell1996, Beach2004}, as useful as it is, is a
numerically ill-posed question and is used with caution for bringing ambiguities
that mask the fine features in the real frequency data. These
difficulties shadow the progress in evaluating the bulk wave
functions and entanglement information for diagnosing interacting
TIs.

In light of the difficult situation for interacting TIs, recently,
a new diagnosis dubbed ``strange correlator" was proposed in
Ref.~\onlinecite{YZYou2014_SC}, which is the matrix element of the
correlation function between two topologically distinct many-body
bulk wave functions in the same Hilbert space. Based on the low-energy
effective Lorentz invariance of the SPT states, the strange
correlator, though it is a purely static quantity, effectively
captures the space-time correlation function at the spatial
interface between two topologically distinct phases. Hence, as
long as there exist symmetry-protected edge states at the spatial
interface, i.e., the two wave functions are topologically distinct,
the strange correlator will diagnose the edge modes, at least for
the noninteracting case. As will become clear in this paper, for
the QMC simulations of interacting TIs, the strange correlator
can diagnose the correlated edge modes as well. Moreover, as the
strange correlator is based on the bulk wave function, there is no
need to explicitly create a real spatial boundary to detect the
gapless edge modes, which, in interacting systems, usually gives
rise to strong finite-size effects. Also, comparing with the
measurements of the entanglement spectrum mentioned above, there is no
need to bifurcate the system for evaluating the strange
correlators. There is also no need to perform imaginary-time
correlation as the strange correlators are static quantities which
avoids the analytical continuation step. These advantages
make the strange correlator physically transparent and technically
much easier to implement in QMC. Yet another advantage of the strange
correlator is that it is generally applicable to both fermionic
and bosonic SPT states, either free or interacting. It is also
applicable to ``crystalline" TI~\cite{LFu2011}, because it
respects all the lattice symmetries (no need for boundaries).

In Ref.~\onlinecite{YZYou2014_SC}, the strange correlator has
been applied only to free fermion topological insulators and some
bosonic SPT states. Later on, in Ref.~\onlinecite{Wierschem2014a,
Wierschem2014b}, it was demonstrated that the strange correlator
can capture the nature of the Haldane phase of 1D spin-1 systems.
It was further shown in Ref.~\onlinecite{Ringel2015} that the
strange correlators of 2D bosonic SPT states can be expressed as
correlation functions of 2D conformal field theory. However, the
most important test, namely, the application of the strange correlator
upon interacting fermion topological insulators to diagnose the
interaction-driven topological phase transition, has never been
performed. Here, by means of large-scale quantum Monte Carlo simulations,
we apply the strange correlator to a very realistic and nonintegrable
model for interacting topological insulators, namely, the Kane-Mele-Hubbard (KMH)
model. We present details on how to evaluate the strange
correlators in determinantal QMC~\cite{AssaadEvertz2008}
simulations for interacting fermionic systems and use it to probe
the topological nature of the interaction-driven quantum phase
transition in the KMH model. Furthermore, the interaction effects
on the helical edge states -- the Luttinger liquid behavior--have
been also clearly captured by the strange correlator measurements in
QMC simulations.

The rest of the paper is organized as follows. In
Sec.~\ref{sec:KMH_SC} the KMH model (\ref{sec:KMH_model}) and
strange correlators in various sectors
(\ref{sec:KMH_QMC}) are introduced, with detailed accounts of
their implementation in QMC simulations. In
Sec.~\ref{sec:results}, the strange correlator in the single-particle
sector (\ref{sec:results_single}) is first demonstrated,
followed by those in two-particle spin and pairing sectors
(\ref{sec:results_two}). In the single-particle sector, the
topological nature of the quantum spin Hall insulator to
antiferromagnetic Mott insulator transition can be clearly seen.
In the two-particle channel, the Luttinger liquid behavior of the
edge modes, is well captured by their corresponding QMC strange
correlator measurements. Section.~\ref{sec:summary} summarizes the
physical and numerical advantages of the strange correlator in
diagnosing interacting TIs and proposes future directions.

\section{Model and numerical method}
\label{sec:KMH_SC}
\subsection{generalized Kane-Mele-Hubbard model}
\label{sec:KMH_model}

\begin{figure}[tp!]
  \centering
  \includegraphics[width=1.0\columnwidth]{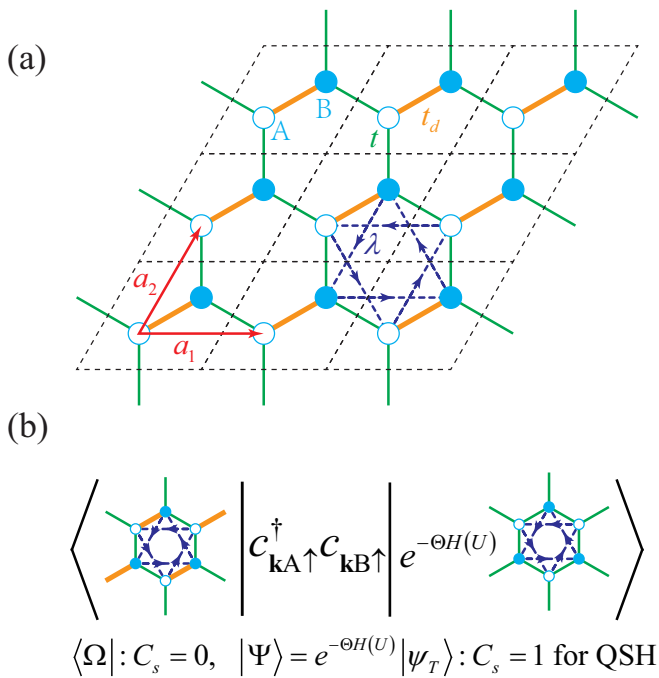}
  \caption{\label{fig:HcLatt}(Color online) (a) Illustration of honeycomb lattice
  and Kane-Mele model. The unit cell of the honeycomb lattice is presented as the dashed
  black parallelogram, it consists of two sublattices, A and B, denoted by the open and
  filled cyan circles. The underlying lattice is spanned by the primitive vectors
  $a_{1}=(\sqrt{3},0)$, $a_2=(\sqrt{3}/2,3/2)$. The green and orange lines represent
  nearest-neighbor hopping $t$ and $t_{d}$ connecting the A and B sublattices.
  The spin-orbital coupling term (complex valued next-nearest-neighbor spin-dependent
  hopping $i\lambda$) connects lattices sites within the same sublattice and is denoted
  as the blue dashed arrows. (b) Schematic plot of $|\Omega\rangle$ and $|\Psi\rangle$ in
  single-particle strange correlator calculation, where $|\Omega\rangle$ is a trivial band
  insulator with spin Chern number $C_{s}=0$, and the detected target state $|\Psi\rangle$
  is the many-body ground wave function of the KMH Hamiltonian in Eq.~\ref{eq:KaneMeleHubbardHamiltonian}
  evaluated in the QMC simulation. It is prepared by applying the projection operator $e^{-\Theta H}$
  onto a noninteracting trivial wave function $|\Psi_{T}\rangle$ (eigenstate of the KM model
  with $U=0$). The topological nature of $|\Psi\rangle$ depends on the interaction strength
  $U/t$. When $U\le U_{c}$ the system is in the QSH insulator phase with spin Chern number
  $C_{s}=1$, but when $U>U_{c}$, the system is the AFMI phase, which spontaneously breaks
  the key symmetry that protects the topological insulator.}
\end{figure}

The generalized KMH model is given by,
\begin{eqnarray}
H_\text{KMH} &=& -\sum_{\langle i,j
\rangle,\sigma}t_{ij}c^{\dagger}_{i\sigma}c_{j\sigma} +
i\lambda\sum_{\langle\!\langle i,j \rangle\!\rangle, \alpha\beta}
v_{ij}\,c^{\dagger}_{i\alpha}\sigma^{z}_{\alpha\beta}c_{j\beta} \nonumber\\
{}&& +\frac{U}{2}\sum_{i}(n_i-1)^2\;.
\label{eq:KaneMeleHubbardHamiltonian}
\end{eqnarray}
Here the first term describes the nearest-neighbor hopping on a
honeycomb lattice. As shown in Fig.~\ref{fig:HcLatt} (a), we set
the nearest-neighbor hopping within one unit cell with amplitude
$t_{d}$, while others are set with amplitude $t$, $t_{d}$ and $t$ can be
different, depending on the context. The second term represents
spin-orbit coupling (SOC)~\cite{Kane2005a, Kane2005b}, which connects
the next-nearest-neighbor sites with a complex (time-reversal
symmetric) hopping with amplitude $\lambda$, and the factor
$v_{ij}=-v_{ji}=\pm1$ depends on the orientation of the two
nearest-neighbor bonds that the electron traverses in going from
site $j$ to $i$. The $\sigma^{z}_{\alpha\beta}$ in the spin-orbit
coupling term furthermore distinguishes the $\uparrow$ and
$\downarrow$ spin states with the opposite next-nearest-neighbor
hopping amplitudes.

Physically, the noninteracting ($U=0$) Kane-Mele (KM)
model~\cite{Kane2005a,Kane2005b} is a spinful model consisting of
two copies of the the Haldane model~\cite{Haldane1988} with
opposite spins. Although the spinless Haldane model breaks the
time-reversal symmetry $Z_2^T$, the spinful KM model is
time-reversal invariant and its ground state is a quantum spin
Hall (QSH) insulator with counter propagating edge modes.
Regarding the symmetries of the KMH model, the model Hamiltonian
in Eq.~\ref{eq:KaneMeleHubbardHamiltonian} has the charge
$U(1)_\text{charge}$ symmetry $c_{i\sigma}\to
e^{\mathrm{i}\theta}c_{i\sigma}$. The spin-rotational symmetry
$SU(2)$ is broken down by the spin-orbit coupling term
$\lambda$ to $U(1)_\text{spin}$, which keeps the spin
rotation only in the $xy$ plane: $c_{i\sigma}\to
e^{\mathrm{i}\sigma\theta}c_{i\sigma}$. So the symmetry group of
the KMH model is $U(1)_\text{charge}\times U(1)_\text{spin}\rtimes
Z_2^T$, which results in a $\mathbb{Z}$ classification. This
implies that the QSH state must be separated from the trivial
vacuum state with gapless edge modes.

In the presence of interaction, the KMH model can be studied by
determinantal QMC simulations~\cite{AssaadEvertz2008,
Hohenadler2011, DZheng2011, Hohenadler2012, Assaad2013, Lang2013,
Hung2013, Hung2014, Meng2014}. At the noninteracting limit,
$U=0$, for any finite $\lambda$, the system is in the QSH state at
zero temperature. Switching on finite but weak interaction
($U/t>0$), the system is adiabatically connected to the
noninteracting QSH. At strong interaction, $U/t$ will drive the
QSH state into an antiferromagnetic ordered Mott insulator (AFMI)
state~\cite{Rachel2010} through a continuous quantum phase
transition at critical point $U_c$ (e.g., at $\lambda=0.1t$, $U_c
\sim 5t$). At the transition, the single-particle gap remains
open but the corresponding spin gap closes~\cite{Hohenadler2012,
Lang2013, Meng2014}. The transition from the QSH to the $xy$ AFMI
has been shown to be consistent with the 3D $XY$ universality
class\cite{Hohenadler2012, Assaad2013, Toldin2015}. As both
$U(1)_\text{spin}$  and $Z_2^T$ symmetries are spontaneously
broken in the AFMI phase, only $U(1)_\text{charge}$ remains.
Meanwhile there is another time-reversal-like symmetry
$Z_{2}^{T^{\prime}}$:
$c_{i\alpha}\to\mathcal{K}\sigma^x_{\alpha\beta}c_{i\beta}$, so
the total remaining symmetry is $U(1)_\text{charge}\rtimes
Z_2^{T^{\prime}}$ with $\mathcal{T}^2=1$. Thus the fermion SPT
classification becomes trivial, and as such, a time-reversal-like
symmetry with $\mathcal{T}^2=1$ does not lead to a Kramers doublet.
This means that if we neglect the Goldstone mode in the bosonic sector,
the AFMI state must belong to the trivial SPT class in the fermionic sector, which can be
smoothly connected to a trivial band insulator [such as a spin
density wave (SDW) insulator]. So there is
no symmetry-protected gapless fermionic edge mode between this
AFMI and a trivial insulator.

Of course, interaction-driven topological phase transitions happen in other models as
well. For example, in the Bernevig-Hughes-Zhang model, dynamical mean-field theory
studies \cite{LeiWang2012, Yoshida2012, Budich2013, Amaricci2015} reveal interesting (first-order) topological phase transitions in the paramagnetic
sector of the solution.

\begin{figure}[tp!]
  \centering
  \includegraphics[width=1.0\columnwidth]{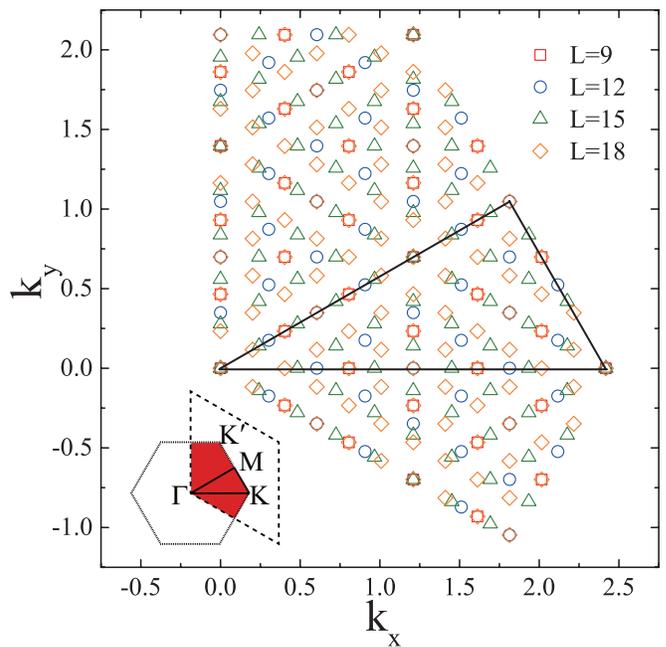}
  \caption{(Color online) Illustration of the $\mathbf{k}$ mesh in the Brillouin zone (BZ)
  of finite-size systems studied in QMC simulations with linear system size $L=9,12,15,18$.
  The black line is the high-symmetry paths $\Gamma \to M \to K \to \Gamma$. As $L$ increases,
  the $\mathbf{k}$ mesh becomes denser. The inset is the hexagon BZ of the honeycomb lattice, and
  the shaded region represents the segment of the BZ shown in the main panel.}
  \label{fig:KPoints_RBZ}
\end{figure}

\subsection{Strange correlator in QMC}
\label{sec:KMH_QMC}

To effectively diagnose the SPT states, the concept of a strange correlator
was proposed in Ref.~\onlinecite{YZYou2014_SC}. It is a correlation function
defined between two many-body wave functions in the same Hilbert space,
\begin{equation}
  C(r,r')=\frac{\langle\Omega|\phi(r)\phi(r')|\Psi\rangle}{\langle\Omega|\Psi\rangle},
\end{equation}
where $|\Omega\rangle$ is a trivial band insulator, and
$|\Psi\rangle$ is the wave function whose topological nature we
would like to diagnose. The physical meaning of $C(r, r^\prime)$
becomes manifest after a space-time rotation~\cite{YZYou2014_SC}:
$|\Psi \rangle$ can be obtained by evolving a generic initial
state from imaginary time $\tau = - \infty$ to $\tau = 0$ with the
parent Hamiltonian of $|\Psi\rangle$, and $|\Omega \rangle$ can be
obtained by evolving a generic final state backward from imaginary
time $\tau = + \infty$ to $\tau = 0$, thus $C(r, r^\prime)$ can be
viewed as a correlation function at the temporal domain wall.
Because most of the topological insulators have an effective
Lorentz invariant description~\footnote{It is well-known that most
topological insulators can be described by Dirac fermions at low
energy, and the bosonic SPT states can be described by either a
nonlinear sigma model field theory~\cite{ZBi2015}, or a
Chern-Simons field theory~\cite{YMLu2012}, both of which have an
effective Lorentz invariance.}, after a space-time rotation $C(r,
r^\prime)$ becomes the space-time correlation at the spatial
interface between $|\Psi\rangle$ and $|\Omega\rangle$, which may
have gapless modes depending on the nature of these two states.

The proposition given in Ref.~\onlinecite{YZYou2014_SC} is that if
$|\Psi\rangle$ is a nontrivial topological insulator (or more
generally, a SPT state) in one or two spatial dimensions, i.e.,
there exists one or more gapless edge modes at the spatial boundary
of $|\Psi\rangle$, then for local operator $\phi(r)$ that
transforms nontrivially under symmetry, $C(r,r')$ will either
develop long-range order (saturate to a {\it constant}) or decay
as a power law in the limit $|r-r'|\to+\infty$, which mimics the
edge states of $|\Psi\rangle$. In the momentum space, this
corresponds to a singularity at a certain symmetric momentum point
$\mathbf{k}_s$: $C_{\mathbf{k}} \sim
1/|\mathbf{k}-\mathbf{k}_s|^\alpha$, if $|\Psi\rangle$ is in a
nontrivial topological insulator phase. Based on the space-time
rotation argument given above, the 2D strange correlator $C(r,
r^\prime)$ should behave very similarly to the (1+1)D correlation
functions at the boundary. For example, if $|\Psi\rangle$ is a
generic {\it noninteracting} 2D TI, and $\phi(r)$ is simply the
electron operator, $i.e.$ $C(r,r^\prime) =
\langle\Omega|c^\dagger(r)
c(r')|\Psi\rangle/\langle\Omega|\Psi\rangle$, then $\alpha = 1$.
The strange correlator has been successfully applied to detect
topological phase transitions in 1D and 2D spin
systems~\cite{YZYou2014_SC, Wierschem2014a, Wierschem2014b}, as
well as in a noninteracting fermionic system~\cite{YZYou2014_SC}.

In our QMC simulations, to detect the correlated QSH phase and
the interaction-driven phase transition in the KMH model, we prepare
$|\Omega\rangle$ as the wave function of
Eq.~\ref{eq:KaneMeleHubbardHamiltonian} with $U=0$ but keep
$t_{d}$ different from $t$. At the noninteracting level, with
finite $\lambda$, $t_{d}/t$ will drive a topological phase
transition between QSH and trivial band insulator at
$t_{d}=2t$~\cite{Lang2013,Hung2014,Meng2014}; therefore
throughout this paper we choose $|\Omega\rangle$ with
$\lambda=0.2t$ and $t_{d}=100t$, which guarantees it is a
topologically trivial band insulator. On the other hand,
$|\Psi\rangle$ is prepared as the ground-state wave function of
the interacting Hamiltonian in Eq.~\ref{eq:KaneMeleHubbardHamiltonian}
with $t_{d}=t$. In the quantum Monte Carlo simulation, it is
prepared as $|\Psi\rangle=e^{-\Theta H}|\Psi_{T}\rangle$,
where $|\Psi_{T}\rangle$ is the wave function of noninteracting
Hamiltonian in Eq.~\ref{eq:KaneMeleHubbardHamiltonian} with $U=0$,
$\lambda=0.2t$ and $t_{d}=t$. The projection operator $e^{-\Theta
H}$ is applied onto $|\Psi_{T}\rangle$ in quantum Monte
Carlo sampling such that when the projection parameter $\Theta$ is
sufficiently large, the QMC ensemble average guarantees
$|\Psi\rangle$ is the ground state of the interacting Hamiltonian
$H$. In most of the simulations, we set $\Theta=50t$.

\begin{figure}[h!]
  \centering
  \includegraphics[width=\columnwidth]{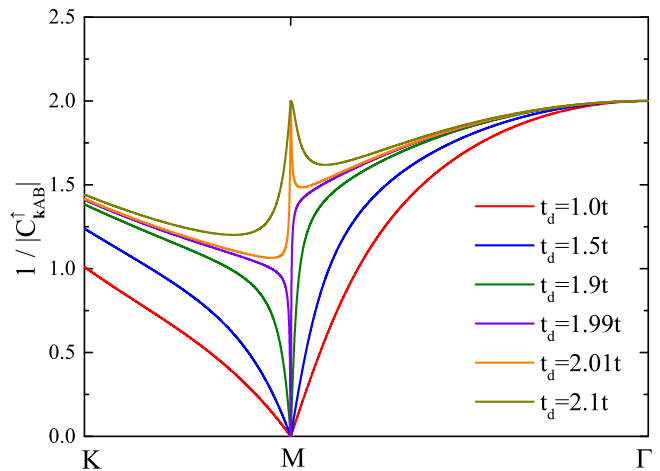}
  \caption{\label{fig:StrCorrNonI}(Color online) $1/|C^{\uparrow}_{\mathbf{k}AB}|$
  as a function of $t_{d}$ in $|\Psi\rangle$. The linear divergence of $|C^{\uparrow}_{\mathbf{k}AB}|$ around the M point
  holds robust until $t_{d} > 2t$. We can use the divergent to nondivergent behavior
  of $|C^{\uparrow}_{\mathbf{k}AB}|$ to determine the critical point precisely in this noninteracting case.}
\end{figure}

\begin{figure}[tp!]
\centering
\includegraphics[width=\columnwidth]{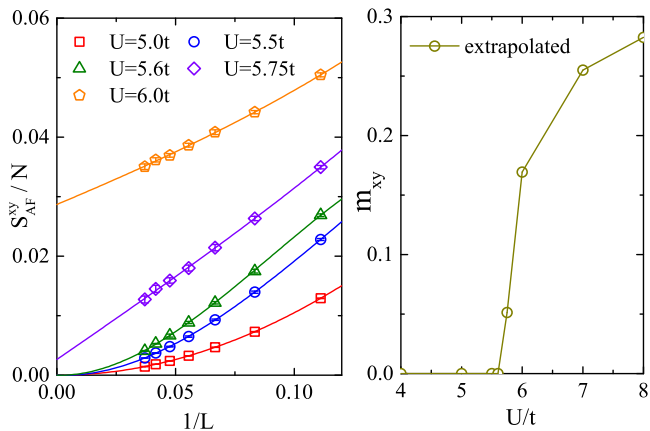}
\caption{\label{fig:StrFct} (Color online) (a). Finite size
scaling of the $xy$ antiferromagnetic structure factor for various
values of $U/t$, with linear system size $L$ going to 27. The
extrapolated values of magnetic moment $m_{xy}$ are plotted in
(b).}
\end{figure}

In this paper, we define the strange correlator in the momentum
space. The strange correlator in the single-particle channel for spin
flavor $\sigma$ is then defined as
\begin{equation}
  C^{\sigma}_{\mathbf{k}AB}=\frac{\langle\Omega|c^{\dagger}_{\mathbf{k}A\sigma}
  c_{\mathbf{k}B\sigma}|\Psi\rangle}{\langle\Omega|\Psi\rangle},
\label{eq:StrCorrSP}
\end{equation}
where
$c^{\dagger}_{\mathbf{k}A\sigma}=\frac{1}{L}\sum_{i}e^{i\mathbf{k}\cdot\mathbf{R}_{i,A}}c^{\dagger}_{i,A,\sigma}$
with $\mathbf{k}$ inside the Brillouin zone (BZ) shown in
Fig.~\ref{fig:KPoints_RBZ}, and $A$, $B$ are the two sublattices
of the honeycomb lattice in Fig.~\ref{fig:HcLatt} (a). The
schematic plot of Fig.~\ref{fig:HcLatt} (b) depicts the idea of
the strange correlator in the KMH model, on the left-hand side, and the
wave function $|\Omega\rangle$ is a trivial band insulator (with
spin Chern number $C_{s}=0$); on the right-hand side, the
projection operator $e^{-\Theta \hat{H}}$ guarantees
$|\Psi\rangle=e^{-\Theta \hat{H}}|\Psi_{T}\rangle$ is the
many-body ground state wave function of the KMH Hamiltonian at certain
$U/t$, although the trial wave function $|\Psi_{T}\rangle$ is
noninteracting (with spin Chern number $C_{s}=1$). In this way,
as we gradually increase the interaction strength $U/t$ in the KMH
Hamiltonian, the nature of $|\Psi\rangle$ will change from QSH at
weak interaction ($U \le U_{c}$) to AFMI at strong interaction
($U>U_{c}$).

We also measure the strange correlator in the spin and Cooper-pair
channels, respectively, as follows:
\begin{eqnarray}
  S^{\pm}_{\mathbf{k}AA}&=&\frac{\langle\Omega|S^{+}_{\mathbf{k}A}S^{-}_{\mathbf{k}A}|\Psi\rangle}{\langle\Omega|\Psi\rangle}, \label{eq:spinstrangecorr}\\
  D_{\mathbf{k}AA}&=&\frac{\langle\Omega|\Delta^{\dagger}_{\mathbf{k}A}\Delta_{\mathbf{k}A}|\Psi\rangle}{\langle\Omega|\Psi\rangle},
\end{eqnarray}
These are two-particle strange correlators in particle-hole and
particle-particle channels, respectively, where
$S^{+}_{\mathbf{k}A}=\frac{1}{L}\sum_{i}e^{i\mathbf{k}\cdot\mathbf{R}_{i,A}}S^{+}_{i,A}$
and
$\Delta^{\dagger}_{\mathbf{k}A}=\frac{1}{L}\sum_{i}e^{i\mathbf{k}\cdot\mathbf{R}_{i,A}}\Delta^{\dagger}_{i,A}$,
with $S^{+}_{i,A}=c^{\dagger}_{i,A,\uparrow}c_{i,A,\downarrow}$
flipping spin in sublattice $A$ of unit cell $i$, and
$\Delta^{\dagger}_{i,A}=c^{\dagger}_{i,A,\uparrow}c^{\dagger}_{i,A,\downarrow}$
creating a Cooper pair of a spin singlet in sublattice $A$ of unit
cell $i$.

Although the magnetic nature of the QSH-to-AFMI transition has
been studied thoroughly~\cite{Hohenadler2012, Meng2014}, here we
find the topological nature of this transition is well captured by
the strange correlators in single- and two-particle sectors. As
will be explained later, the QMC computation of the strange
correlator is more efficient and robust than the QMC simulations
with either open boundary conditions (OBCs) to directly probe the
edge modes~\cite{Hohenadler2011, DZheng2011}, or measurements of the
entanglement spectrum, where one has to bifurcate the already small
finite-size system and analytically continue the imaginary-time
data~\cite{Assaad2014, Assaad2015}.

\section{NUMERICAL RESULTS AND DISCUSSIONS}
\label{sec:results}
\subsection{Single-particle strange correlator}
\label{sec:results_single}

We first apply the single-particle strange correlator to detect
the topological phase transition driven by $t_{d}$ at the
noninteracting limit. In Fig.~\ref{fig:StrCorrNonI}, we set $U=0$
but gradually increase $t_{d}$ in $|\Psi\rangle$. One clearly sees
that when $t_{d}<2t$, the strange correlator
$|C^{\uparrow}_{\mathbf{k}AB}|$ is linearly divergent at one M
point in the Brillouin zone, which is consistent with the
prediction in Ref.~\onlinecite{YZYou2014_SC}. When $t_{d}>2t$,
both $\bra{\Omega(t_{d}=100t)}$ and $\ket{\Psi(t_{d}>2t}$ become
a topological trivial state ($C_{s}=0$), the divergence of
$|C^{\uparrow}_{\mathbf{k}AB}|$ is removed.
$1/|C^{\uparrow}_{\mathbf{k}AB}|$ shows an upturn behavior around
$\mathbf{k}_{M}$.

Before we move on to the strange correlator in the interacting case,
we first look at the phase transition from QSH to AFMI from the
magnetic perspective. Figure.~\ref{fig:StrFct} (a) shows the $1/L$
extrapolation of the antiferromagnetic structure factor:
\begin{equation}
S^{xy}_{AF}=\frac{1}{4L^2}\sum_{\langle i,j \rangle}\sum_{\alpha=A,B}
\langle S^{+}_{i,\alpha}S^{-}_{j,\alpha} +
S^{-}_{i,\alpha}S^{+}_{j,\alpha} \rangle
\end{equation}
for various values of $U/t$. $\langle \cdots \rangle$ indicates the QMC
average with $|\Psi\rangle$ on both sides of the observable;
hence $S^{xy}_{AF}$ is not measured as strange correlator but as
a regular QMC correlator. From the extrapolated values of
$L\to\infty$, one can see the $xy$ antiferromagnetic order sets in
at around $U_{c}\approx 5.7t$ which is consistent with previous QMC
results~\cite{Assaad2013, Toldin2015}. The corresponding magnetic
moment is obtained as $m_{xy}=\sqrt{S^{xy}_{AF}/L^2}$, and its
value is plotted as a function of $U/t$ in Fig.~\ref{fig:StrFct}
(b). The appearance of magnetic long-range order breaks the
time-reversal symmetry and destroys the bulk topological state.
Previous theoretical and numerical studies show that the counter
propagating edge modes in the QSH phase are expected to become
gapped exactly at the point where long-range magnetic order in the
bulk breaks time-reversal symmetry~\cite{Assaad2013_JPCM}.

\begin{figure}[tp!]
  \centering
  \includegraphics[width=\columnwidth]{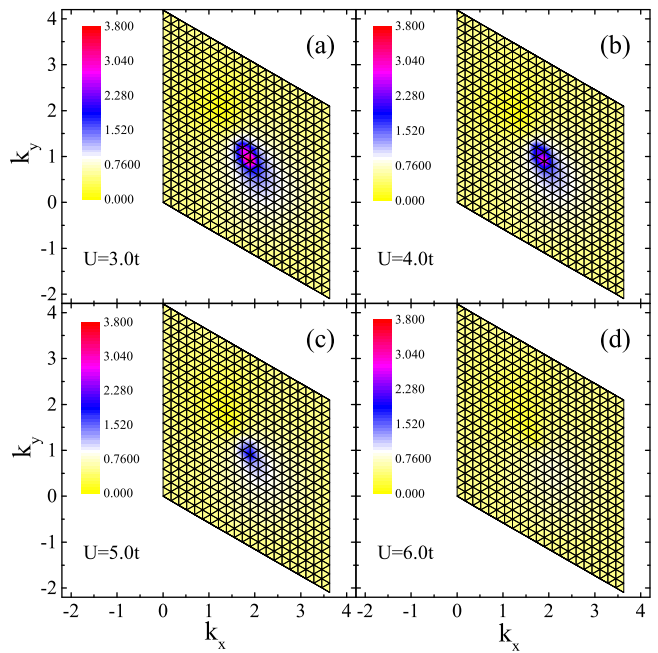}
  \caption{\label{fig:StrCorrU3D} (Color online) The contour plot of
  single-particle strange correlator $|C_{\mathbf{k}AB}^{\uparrow}|$
  with increasing Hubbard interaction $U/t$. The finite system size
  used here is $L=21$. The $\mathbf{k}$-space area in the four
  panels is the same as the dashed region in the inset of
  Fig.~\ref{fig:KPoints_RBZ}, which is the whole BZ.}
  \label{fig:OccST3D}
\end{figure}

After determining the critical $U_c$ from a magnetic perspective, we
monitor the single-particle strange correlator
$|C^{\uparrow}_{\mathbf{k}AB}|$ as a function of $U/t$ in the
whole Brillouin zone. For a global view, Fig.~\ref{fig:StrCorrU3D}
shows the contour plot of the strange correlator
$|C_{\mathbf{k}AB}^{\uparrow}|$ with increasing $U/t$ for a fixed
system size $L=21$. When $U$ is small, there is a clear
singularity at one and only one of the time-reversal-symmetric M
points. In the thermodynamic limit, the single-particle strange correlator
is still divergent at one M point in Figs.~\ref{fig:OccST3D} (a-c).
When $U>U_{c}$ [see Fig.~\ref{fig:OccST3D} (d)], there will be no
divergence in the single-particle strange correlator.

\begin{figure}[ht!]
  \centering
  \includegraphics[width=\textwidth]{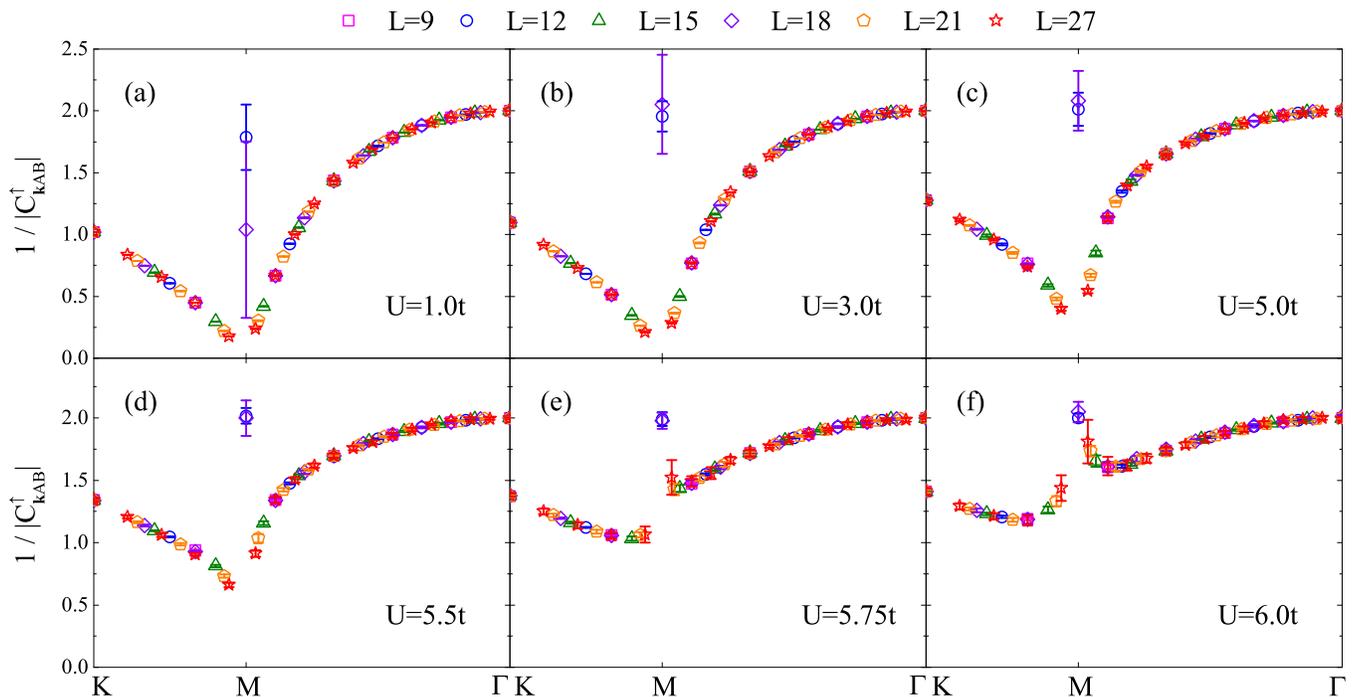}
  \caption{\label{fig:StrCorrInvKMG} (Color online) The inverse amplitude
  of single-particle strange correlator $1/|C_{\mathbf{k}AB}^{\uparrow}|$
  along the high-symmetry path for various $U/t$ and system sizes. When the
  interaction $U/t\le 5.5t$, see (a-d), there is a divergent tendency in $|C_{\mathbf{k}AB}^{\uparrow}|$
  around the M point. However, the divergent exponent is reduced due to the
  correlation effects according to the helical Luttinger liquid theory
  (see the main text). In the AFMI regime (e-f), $1/|C_{\mathbf{k}AB}^{\uparrow}|$
  shows upturn behavior around the M point; there is no divergence at all.}
\end{figure}

A careful analysis of $1/|C_{\mathbf{k}AB}^{\uparrow}|$ along the
high-symmetry path K$\rightarrow$M$\rightarrow$G is shown in
Fig.~\ref{fig:StrCorrInvKMG}. In Figs.~\ref{fig:StrCorrInvKMG}
(a-d), the single-particle strange correlator still shows
divergent tendency at the M point with the finite system size up to
$L=27$. When $U>U_{c}$ [see Figs.~\ref{fig:StrCorrInvKMG} (e) and ~\ref{fig:StrCorrInvKMG}(f)],
where the detected target state $\ket{\Psi}$ becomes
topologically trivial, we clearly see the upturn behavior
around the M point in $1/|C_{\mathbf{k}AB}^{\uparrow}|$.

To give a better understanding of the results in
Fig.~\ref{fig:StrCorrInvKMG}, we turn to the helical Luttinger
liquid theory. Based on the space-time rotation interpretation of
the strange correlator, we can analyze the single-particle strange
correlator using the helical Luttinger liquid theory at the (1+1)D
boundary~\cite{CXu2006, CWu2006, Chung2014, Hohenadler2013},
according to which the real-space strange correlator in the
single-particle sector scales as
\begin{equation}
C^{\sigma}_{\mathbf{r}AB} \sim \mathbf{r}^{-g/2-1/2g},
\end{equation}
where $g$ is the Luttinger parameter related to $U/t$, $g\in [0,1]$. After Fourier transform to the momentum space,
it becomes
\begin{equation}
C^{\sigma}_{\mathbf{k}AB} \sim \tilde{\mathbf{k}}^{g/2+1/2g-2},
\label{eq:OccSTABK}
\end{equation}
where $\tilde{\mathbf{k}}=|\mathbf{k}-\mathbf{k}_{M}|$. Unlike the
noninteracting case, the single-particle strange correlator in the
momentum space may actually stop diverging before the QSH to AFMI
transition point. To see this point more clearly, the critical
$g_{c}$ can be solved from the equation $g/2+1/2g-2=0$, which
gives $g_c=2-\sqrt{3}\approx 0.268$. If $g<g_{c}$, there would be
no divergent behavior around the M point in the momentum space of
the single-particle strange correlator, although the real-space
strange correlator still obeys a power-law decay. For $g>g_{c}$,
the power-law divergent behavior of the single-particle strange
correlator around the M point clearly signifies that the
interacting QSH phase and the trivial band insulator belongs to
distinct SPT phases, and the two states must be separated by
gapless fermion edge modes when they are adjacent in the space.
From the data in Fig.~\ref{fig:StrCorrInvKMG}, the divergent
behavior persists up to $U=5.5t$, which is very close to the
quantum critical point extracted from previous QMC simulations.
From Fig.~\ref{fig:StrCorrInvKMG} here and
Fig.~\ref{fig:Luttingerg} in Sec.~\ref{sec:results_two}, we can
see that the divergent exponent of the single-particle strange
correlator is reduced by the interaction, which cannot be captured
by the noninteracting topological phase transition in
Fig.~\ref{fig:StrCorrNonI} and is clearly beyond the mean-field
level.

\onecolumngrid

\twocolumngrid

We notice that the data points exactly at
$\mathbf{k}=\mathbf{k}_{M}$ in Figs.~\ref{fig:StrCorrInvKMG} (a-d)
suddenly jump up and have larger error bars. This is unphysical,
and we will discuss the behavior of $C_{\mathbf{k}AB}^{\uparrow}$
in the presence of small antiferromagnetic (AF) order $\Delta_\text{SDW}$ around the M
point in a mean-field context in Appendix~\ref{sec:appendix_a},
where this unphysical singularity at $\mathbf{k}=\mathbf{k}_{M}$
will be understood.

We want to stress that based on the Luttinger liquid theory the
single-particle strange correlator, and equivalently, the single
particle Green's function at the physical edge of the system,
always follow a power-law decay before the system develops a true
long-range order in the bulk. This is mainly because when the
bulk is fully gapped, all the low-energy physics occur at the
boundary of the system. Then, based on the Mermin-Wagner
theorem~\cite{MW1966}, continuous symmetries cannot be
spontaneously broken in a $(1+1)$D system, and without a true long-range
correlation of magnetic or superconductor order parameter,
the fermions at the boundary should remain gapless (though still
strongly interacting).

The technical advantage of the strange correlator in QMC over other
numerical diagnoses of interacting TIs is manifestly presented,
i.e., we have performed simulations on finite-size systems with
periodic boundary conditions (PBCs) yet still are able to extract
information on the edge modes, which, in the past, could only be
obtained with systems with OBCs~\cite{Hohenadler2011, DZheng2011}.
It is well known that QMC simulations with OBCs suffer from greater
finite-size effects, but apparently the strange correlator avoids this
difficulty. Moreover, direct probe of edge modes with OBCs requires
an analytical continuation of the imaginary-time Green's function, i.e.,
from $G(\mathbf{k},\tau)$ to $A(\mathbf{k},\omega)$, and that
usually renders ambiguity in the real-frequency data. However,
with the strange correlator, we need to measure only the static (equal
time) single-particle Green's function in the PBC system, which is the
easiest and most reliable observables in the QMC simulations.
Third, as mentioned in the Introduction, in comparison with
measurements of the entanglement spectrum to detect the
interaction-driven topological transition~\cite{Grover2013,
Assaad2014, Assaad2015}, the strange correlator is also physically
more transparent and technically more robust, as in the
entanglement spectrum measurements one has to bifurcate the
already small finite-size system and analytically continue the
imaginary-time data, whereas in the strange correlator both
problems are avoided. Hence, at the technical level, to the best
of our knowledge, the strange correlator is indeed the easiest
diagnosis of the topological states and the topological quantum
phase transition in interacting systems.

\onecolumngrid

\begin{figure}[ht!]
  \begin{center}
  \includegraphics[width=\columnwidth]{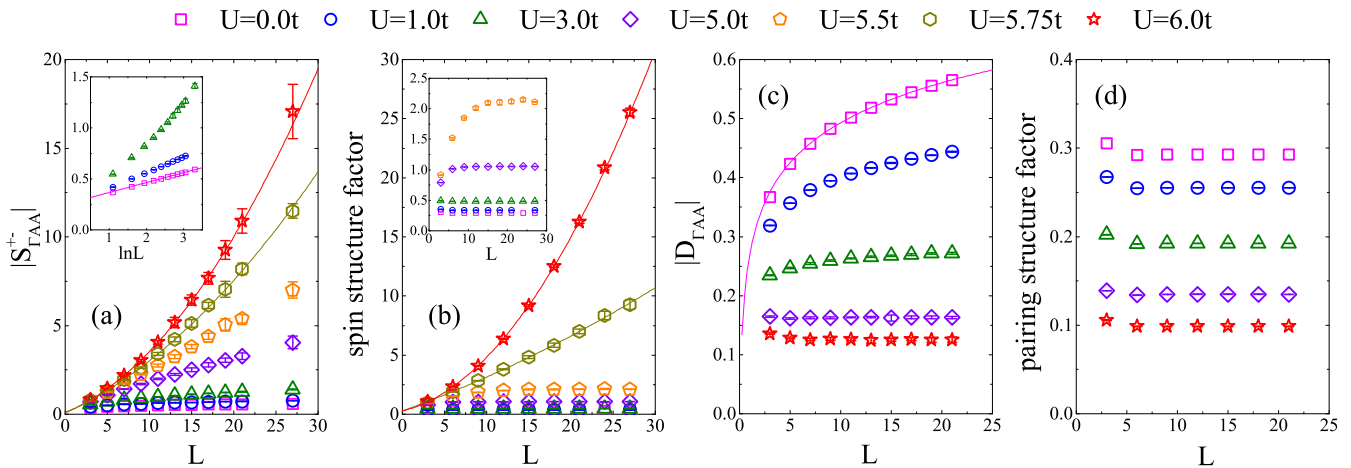}
  \caption{(Color online) (a,c) Spin and pairing strange
  correlators $|S^{\pm}_{\Gamma AA}|$ and $|D_{\Gamma AA}|$ as a
  function of system size $L$ for various $U/t$. The inset of (a) is
  a zoomin at small $U/t$, with a logarithmic fit (magenta solid
  line) of the data at $U=0$ according to
  Eq.~\ref{eq:spinstrangecorrlog}. The dark yellow and red solid lines in the main
  panel of (a) are power-law fits according to
  Eq.~\ref{eq:spinstrangecorrpower2} at $U=5.75t (\sim U_{c})$ and $U=6t$. The logarithmic fit
  (magenta solid line) in (c) also follows
  Eq.~\ref{eq:pairingstrangecorrlog} at $U=0$. (b,d) Spin and
  pairing regular correlation functions $\langle \Psi| S^+_k S^-_k |\Psi \rangle$ and
  $\langle\Psi | \Delta^\dagger_k \Delta_k |\Psi\rangle$
  as a function of $L$ for various $U/t$. The dark yellow and red solid lines in (b)
  are power-law fits according to
  Eq.~\ref{eq:spinstrangecorrpower2} at $U=5.75t(\sim U_{c})$ and $U=6t$. The inset of (b) is
  a zoomin of the regular spin structure factor at small $U/t$,
  showing that they are independent of $L$ when $U<U_c$.}
\label{fig:spin_pairing}
\end{center}
\end{figure}

\twocolumngrid

\subsection{Two-particle strange correlators}
\label{sec:results_two}

In this section, we discuss the QMC results on strange correlators
in the two-particle sector, i.e., the spin and pairing strange
correlators in the presence of interaction.

Again, based on the space-time rotation interpretation of the
strange correlator, we can likewise analyze the spin and pairing
strange correlators using the helical Luttinger liquid theory at the
$(1+1)$D boundary~\cite{CXu2006, CWu2006, Chung2014,
Assaad2013_JPCM}. According to the theory, the real-space strange
correlator in the spin and pairing sectors scale as
\begin{eqnarray}
S^{\pm}_{\mathbf{r}AA} &\sim& \mathbf{r}^{-2g},\\
D_{\mathbf{r}AA} &\sim& \mathbf{r}^{-2/g},
\end{eqnarray}
where $g$ is the Luttinger parameter. After Fourier transform to
the momentum space, they become
\begin{eqnarray}
S^{\pm}_{\mathbf{k}AA} &\sim& \tilde{\mathbf{k}}^{2g-2}, \quad (\sim L^{2-2g} \; \text{at} \; \mathbf{k}=\Gamma), \label{eq:spinstrangecorrL}\\
D_{\mathbf{k}AA} &\sim& \tilde{\mathbf{k}}^{2/g-2}, \quad (\sim L^{2-2/g}
\; \text{at} \; \mathbf{k}=\Gamma). \label{eq:pairingstrangecorrL}
\end{eqnarray}
where $\tilde{\mathbf{k}}=|\mathbf{k}-\mathbf{k}_{\Gamma}|$. In
the noninteracting limit ($U=0$) $g = 1$, and as we increase $U/t$
towards $U_{c}$, $g$ will become smaller and smaller, eventually
vanish at the transition point.

To better understand the behavior in each limit, let us start with
$g = 1$ ($U=0$), and we have
\begin{eqnarray}
S^{\pm}_{\mathbf{k}AA} &\sim& \tilde{\mathbf{k}}^{0} \sim \ln(\mathbf{k}),
\quad (\sim \ln(L) \; \text{at} \; \mathbf{k}=\Gamma),
\label{eq:spinstrangecorrlog}\\
D_{\mathbf{k}AA} &\sim& \tilde{\mathbf{k}}^{0} \sim \ln(\mathbf{k}), \quad
(\sim \ln(L) \; \text{at} \; \mathbf{k}=\Gamma).
\label{eq:pairingstrangecorrlog}
\end{eqnarray}
Such a logarithmic growth in $L$ fits our calculated data in
Figs.~\ref{fig:spin_pairing} (a) and ~\ref{fig:spin_pairing} (c) for the $U=0$ cases very
well. The logarithmic growth is in strong contrast to the regular
spin and pairing correlators, as shown in
Figs.~\ref{fig:spin_pairing} (b) and ~\ref{fig:spin_pairing} (d), which, at $U=0$, are
independent of system size $L$, meaning both spin and pairing
correlations are exponentially short ranged in real space,
corresponding to the QSH insulator with a bulk gap.

On the other hand, near the QSH-to-AFMI transition point, $g = 0$
($U \sim U_{c}$), we have
\begin{eqnarray}
S^{\pm}_{\mathbf{k}AA} &\sim& \tilde{\mathbf{k}}^{-2}, \quad (\sim L^{2}
\; \text{at} \; \mathbf{k}=\Gamma),
\label{eq:spinstrangecorrpower2}\\
D_{\mathbf{k}AA} &\sim& \tilde{\mathbf{k}}^{\infty}, \quad (\sim
L^{-\infty} \sim e^{-L} \; \text{at} \; \mathbf{k}=\Gamma).
\label{eq:pairingstrangecorr}
\end{eqnarray}
As we can see, the calculated data in Fig.~\ref{fig:spin_pairing} (a) at
$U=6t$ indeed diverge as $L^{2}$ in the thermodynamic limit.
More interestingly, such a divergence is the same as the one
shown by the regular spin correlator inside the AFMI phase,
as shown in Fig.~\ref{fig:spin_pairing} (b) at $U=6t$. This is
because when $U=6t$, the ground state wavefunction $|\Psi\rangle$
in Eq.~\ref{eq:spinstrangecorr} is already in the AFMI phase; the
spin strange correlator is then similar to the spin regular
correlator, because both of them pick up the long-range spin-spin
correlation. In Fig.~\ref{fig:spin_pairing} (c) at $U=6t$, the
pairing strange correlator decays exponentially to a constant,
also consists with the prediction in
Eq.~\ref{eq:pairingstrangecorr}.

\begin{figure}[htp!]
\begin{center}
  \includegraphics[width=\columnwidth]{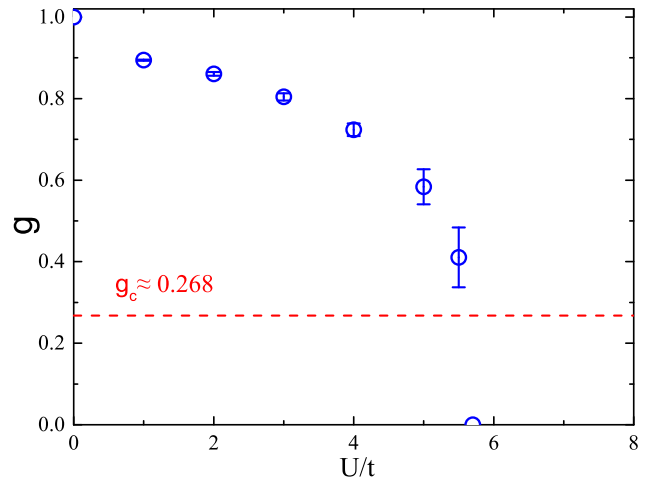}
  \caption{(color online) The Luttinger parameter $g$ extracted from
  the spin strange correlator in Fig.~\ref{fig:spin_pairing} (a)
  following Eq.~\ref{eq:spinstrangecorrL}. Below the critical value $g_{c}$,
  the single-particle strange correlator in the momentum space is no longer
  divergent at M point.} \label{fig:Luttingerg}
\end{center}
\end{figure}

Between the limits of $g=1$ ($U=0$) and $g=0$ ($U=U_{c}$), we can
fit the spin and pairing strange correlator data in
Figs.~\ref{fig:spin_pairing} (a) and ~\ref{fig:spin_pairing} (c) with the Luttinger liquid
theory prediction in Eqs.~\ref{eq:spinstrangecorrL}  and
\ref{eq:pairingstrangecorrL}, to extract the Luttinger parameter
$g$. The extracted $g$ values as a function of $U/t$ are shown in
Fig.~\ref{fig:Luttingerg}. One can see that $g$ continuously
decreases from 1 to 0, which accounts for the increasing
electron-electron correlation. The dashed line in
Fig.~\ref{fig:Luttingerg} highlights the $g_c$ smaller than which
the single-particle strange correlator stops diverging, as
discussed in Sec.~\ref{sec:results_single}.

\section{SUMMARY AND OUTLOOK}
\label{sec:summary}

In summary, we have employed large-scale QMC simulations to study
the single-and two-particle strange correlators in a realistic
model for interacting topological insulators. We demonstrate that
the interaction-driven topological-to-trivial quantum phase
transition can be well captured by the strange correlators.
Although larger system sizes might be needed for detailed
information very close to the critical point, our results show
that the strange correlator is a powerful and promising tool to
diagnose the topological insulator with interaction.

The technical advantages of the strange correlator in numerical studies
(especially QMC simulations) on interacting fermionic and bosonic
SPT states are obvious. As one needs to measure only static
correlations in the bulk system, there is no need to apply OBCs to
actually probe the spatial edges, no need to apply analytical
continuation to access real-frequency data, and no need to
bifurcate the already small finite-size systems for entanglement
measurements. In short, the strange correlator is much easier to
implement and robust in practical numerical performance.

As for future applications, the QSH insulator discussed in our
work has a full spin $S^z$ conservation, which has a $\mathbb{Z}$
classification instead of a $\mathbb{Z}_2$ classification for the
cases with time-reversal symmetry but no $S^z$ conservation. In
Ref.~\onlinecite{YZYou2014_SC} the strange correlator was tested
for a noninteracting QSH insulator with a sizable Rashba
spin-orbit coupling, which does have a $\mathbb{Z}_2$
classification. We expect that the same strange correlator is still
applicable to the interacting QSH insulator with Rashba spin-orbit
coupling as well, except now that the two electron operators in
the strange correlator equation.~\ref{eq:StrCorrSP} do not have to have
the same spin, since the spin conservation is broken by the Rashba
term.

As we mentioned in the Introduction, in all dimensions interaction
can change or reduce the classification of some topological insulators,
for example, interaction may trivialize some topological insulators that
are nontrivial in the noninteracting limit. This means that in
this case the strange correlator should be power-law or long range
correlated without interaction, but becomes short-ranged due to
interaction, possibly even {\it without} going through any bulk
phase transition. We will leave this to future study.

\begin{acknowledgments}

The numerical calculations were carried out at the Physical
Laboratory of High Performance Computing in RUC as well as the
National Supercomputer Center in Guangzhou on the Tianhe-2 platform.
H.Q.W., Y.Y.H. and Z.Y.L. acknowledge support from the National
Natural Science Foundation of China (Grant Nos. 11474356 and
91421304) and the National Program for Basic Research of MOST of China
(Grant No. 2011CBA00112). C.X. and Y.Z.Y. are supported by the David
and Lucile Packard Foundation and NSF Grant No. DMR-1151208. Z.Y.M.
is supported by the National Thousand-Young-Talents Program of
China.

\end{acknowledgments}

\appendix

\section{A Mean-Field Calculation of the Strange Correlator}
\label{sec:appendix_a}

To facilitate the understanding of the behavior of the single-particle
strange correlator in the QSH insulator to $xy$ AFMI  transition
in the KMH model, below, we also provide a mean-field-level
calculation of $1/|C_{\mathbf{k}AB}^{\uparrow}|$ by introducing
the SDW order parameter, $\Delta_\text{SDW}$.
The mean-field Hamiltonian can be written as
\begin{equation}
\begin{split}
H_\text{MF}=&-\sum_{\langle ij\rangle,\sigma}t_{ij}c_{i\sigma}^\dagger c_{j\sigma}+i\lambda\sum_{\langle\!\langle ij\rangle\!\rangle,\alpha,\beta}v_{ij}c_{i\alpha}^\dagger\sigma^z_{\alpha\beta}c_{j\beta}\\
&-\Delta_\text{SDW}\sum_{i,\alpha,\beta}(-)^ic_{i\alpha}^\dagger
\sigma^x_{\alpha\beta}c_{i\beta},
\end{split}
\end{equation}
where $\Delta_\text{SDW}$ is the SDW gap. Here we set $\lambda=0.2t$.
If $t_d=t$ and $\Delta_\text{SDW}=0$, $H_\text{MF}$ describes the QSH
insulator. The trivial band insulator can be obtained by tuning $t_d>2t$.
The strong interacting AFMI can be phenomenologically modeled by a
finite $\Delta_\text{SDW}$ term in the mean-field theory, which
breaks the spin $U(1)$ symmetry and describes the spin ordered
antiferromagnetic state. Here we assume the $xy$ spin order lies
in the spin-$x$ direction.

\begin{figure}
\begin{center}
\includegraphics[width=\columnwidth]{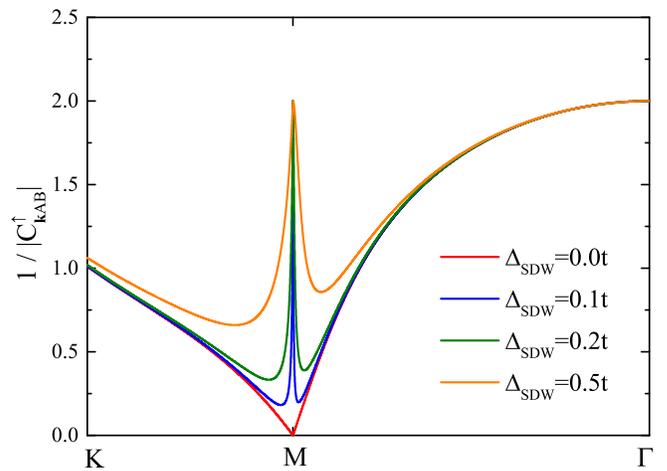}
\caption{The inverse strange correlator
$1/|C_{\mathbf{k}AB}^{\uparrow}|$ along the high-symmetry path
in the mean-field theory. The state $|\Psi\rangle$ is replaced by
an SDW insulator controlled by $\Delta_\text{SDW}$.}
\label{fig:StrCorrInvMF}
\end{center}
\end{figure}

We calculated the inverse strange correlator
$1/|C_{\mathbf{k}AB}^{\uparrow}|$ with the state $|\Psi\rangle$
tuned by the mean-field parameter $\Delta_\text{SDW}$. The result
is shown in Fig.\,\ref{fig:StrCorrInvMF}. When
$\Delta_\text{SDW}=0$, $|\Psi\rangle$ is the QSH state, the
inverse strange correlator
$1/|C_{\mathbf{k}AB}^{\uparrow}|\sim|\mathbf{k}-\mathbf{k}_M|$
follows the linear behavior around the M point, which implies the
power law behavior of the strange correlator
$|C_{\mathbf{k}AB}^{\uparrow}|\sim|\mathbf{k}-\mathbf{k}_M|^{-\alpha}$
with $\alpha=1$.  However, beyond the mean-field theory, the
interaction can modify this power $\alpha$, so the strange
correlator can deviates from the $\alpha=1$ behavior in the
momentum space, as shown in Eq.~\ref{eq:OccSTABK} and
Fig.~\ref{fig:Luttingerg} in the main text. But the power-law
behavior of the strange correlator in the real space is still
expected to survive in the whole QSH phase.

As $\Delta_\text{SDW}$ is turned on,
$1/|C_{\mathbf{k}AB}^{\uparrow}|$ will be lifted from zero at the
M point and replaced by a small peak. The stronger SDW order will
leads to the earlier upturn of the curve as approaching to the M
point. The upturn behavior around the M point can be described by
\begin{equation}\label{eq: C small k}
C_{\mathbf{k}AB}^\uparrow=\frac{(k+im_\Omega)(\Delta_\text{SDW}^2m_\Omega
+k(k-im_\Psi)(m_\Omega+m_\Psi))}{2(\Delta_\text{SDW}^2m_\Omega^2+k^2(m_\Omega+m_\Psi)^2)},
\end{equation}
where $k=v_F|\mathbf{k}-\mathbf{k}_M|$ is the small momentum
deviation from the M point, and $m_\Omega$ and $m_\Psi$ are,
respectively, the single-particle mass gaps in the trivial state
$\ket{\Omega}$ and the QSH state $\ket{\Psi}$. Equation.~\ref{eq: C
small k} is derived by a small momentum expansion around the M
point. As can be seen from the denominator, the power law
divergence of the strange correlator
$|C_{\mathbf{k}AB}^{\uparrow}|$ (as $k\to0$) will be removed once
the SDW order $\Delta_\text{SDW}$ sets in at the topological
transition to the AFMI phase. Additionally, according to
Eq.~\ref{eq: C small k}, the limit of $\Delta_\text{SDW} \to 0$
and the limit of $\mathbf{k}\to\mathbf{k}_M$ do not commute:
\begin{equation}
\begin{split}
\lim_{k\to0} C_{\mathbf{k}AB}^{\uparrow} &= \frac{i}{2},\\
\lim_{\Delta_\text{SDW}\to0} C_{\mathbf{k}AB}^{\uparrow}
&=\frac{(k+im_\Omega)(k-im_\Omega)}{k(m_\Omega+m_\Psi)}\sim\frac{1}{k},
\end{split}
\end{equation}
If one takes $\Delta_\text{SDW}\to 0$ first, then the strange
correlator $C_{\mathbf{k}AB}^{\uparrow}$ indeed follows the $1/k$
power-law behavior as expected on the mean-field level. However,
in our QMC simulation, we take $k\to0$ first due to the presence
of the AF fluctuations as a result of the finite-size effect, so
the strange correlator $C_{\mathbf{k}AB}^{\uparrow}$ approaches
another limit $i/2$, which is not divergent. It is this
non-commutative limit that makes $|C_{\mathbf{k}AB}^{\uparrow}|$
ill-defined at the M point and the data right at the M point
meaningless in Fig.~\ref{fig:StrCorrNonI} (a-d). Only when the
interaction becomes sufficiently strong (the Luttinger parameter
$g<g_{c}$), the single-particle strange correlator will no longer diverge,
does the data of $1/|C_{\mathbf{k}AB}^{\uparrow}|$ at
$\mathbf{k}=M$ becomes meaningful.

\bibliography{StrCorrBib}

\end{document}